\documentclass{iopart}
\usepackage[dvips]{graphicx}
\usepackage{latexsym}
\begin{document}
\jl{3}
\title[Breakdown of the magnetization plateau of a spin chain]
{Breakdown of the magnetization plateau 
of the $S=1/2$ ferromagnetic-ferromagnetic-antiferromagnetic trimerized 
spin chain with anisotropy}
\author{Atsuhiro Kitazawa\dag and Kiyomi Okamoto\ddag}
\address{\dag\ Department of Physics, Kyushu University 33, 
Fukuoka 812-8581, Japan}
\address{\ddag\ Department of Physics, Tokyo Institute of Technology, 
Oh-okayama, Meguro-ku, Tokyo 152-8551, Japan}

\begin{abstract}
We study the breakdown of the magnetization plateau 
at the magnetization $M=M_{S}/3$ ($M_{S}$ is the saturation magnetization) 
of the $S=1/2$ anisotropic 
spin chain with ferromagnetic-ferromagnetic-antiferromagnetic interactions. 
We consider the model with the isotropic ferromagnetic (trimer) 
coupling $J_{F}$, 
and anisotropic antiferromagnetic coupling 
($J_{x}=J_{y}=J_{AF}$ and $J_{z}=\Delta J_{AF}$). 
For the limit of large $\gamma\equiv J_{F}/J_{AF}$, 
the model is equivalent to the $S=3/2$ 
$XXZ$ chain with the exchange anisotropy $\Delta$. 
There is a phase transition between the plateau (small-$\gamma$) 
and the no-plateau (large-$\gamma$) regions. 
This phase transition is of the Berezinskii-Kosterlitz-Thouless type, 
and we determine the phase boundary from the numerical diagonalization data. 
For $\Delta=1$, in particular, 
the phase transition between the plateau and the no-plateau 
regions occurs at the point $\gamma_{c}=15.4$. 
\end{abstract}

\section{Introduction}
Recently, there has been considerable interest 
in the magnetization processes of one-dimensional quantum spin systems. 
In particular, magnetization plateaues have been observed in experimental and 
theoretical works, for cases such as 
the $S=1/2$ ferromagnetic-ferromagnetic-antiferromagnetic 
trimerized spin chain \cite{Hida,Okamoto}, 
the $S=1$ spin chain with bond alternation (and single ion  
anisotropy) \cite{Tonegawa1,Totsukak,Narumi,Hagiwara}, 
the $S=3/2$ spin chain with single ion 
anisotropy \cite{Oshikawa,Sakai}, $N$-leg spin ladders \cite{Cabra,Cabra2}, 
the $S=1/2$ zigzag chain with bond alternation 
\cite{Tonegawa2,Totsuka}, 
and the alternating-spin Heisenberg chain 
with $S=1/2$ and $1$ \cite{Kuramoto,Yamamoto}. 
This phenomenon stems from the strong quantum fluctuation due to the 
low dimensionality. 
Extending the Lieb-Schultz-Mattis theorem \cite{LSM}, 
Oshikawa, Yamanaka, and Affleck \cite{Oshikawa} 
gave a necessary condition for the appearance of the magnetization plateau 
as a fractional quantized form, $p(S-\langle m\rangle)=\mbox{integer}$, 
where $p$ is the periodicity of the magnetic ground states 
in the thermodynamic limit, $S$ is the magnitude of the spin, 
and $\langle m\rangle$ is the magnetization per site. 
Because this is a necessary condition, 
it depends on the details of the model whether the magnetization plateau 
exists or not, even if their condition is satisfied. 
In fact, there are no magnetization plateau for simple antiferromagnetic 
Heisenberg spin chains (except for zero magnetization (Haldane gap)) 
\cite{Bonner}. 

In this paper, we study the $S=1/2$ trimerized $XXZ$ spin chains 
in a magnetic field $h$:
\begin{eqnarray}
\fl  H &=& \sum_{j=1}^{L}\left[
  -J_{F}(H_{3j-2,3j-1}(1)+H_{3j-1,3j}(1))
  +J_{AF} H_{3j,3j+1}(\Delta)\right] - hM
\label{eq:ham1} \\
\fl  &=& J_{AF}\sum_{j=1}^{L}\left[
  -\gamma(H_{3j-2,3j-1}(1)+H_{3j-1,3j}(1))
  +H_{3j,3j+1}(\Delta)\right] - hM
\nonumber
\end{eqnarray}
where 
\begin{equation}
 H_{j,k}(\Delta) = S^{x}_{j}S^{x}_{k}+S^{y}_{j}S^{y}_{k}
    +\Delta S^{z}_{j}S^{z}_{k}, 
\end{equation}
$S^{\alpha}_{j}$ ($\alpha=x,y,z$) is the spin-$1/2$ operator 
at the $j$th site, 
$J_{F}=\gamma J_{AF}$, $J_{AF} > 0$, 
and 
\[
M=\sum_{k=1}^{3L}S_{k}^{z}. 
\]
We consider the parameter region $-1<\Delta \leq 1$. 
For $\Delta = 1$, this is a model of 3CuCl$_{2}\cdot$2dioxane, for which 
Ajiro {\it et al} \cite{Ajiro} measured the magnetization process. 
Hida studied this model using the numerical diagonalization 
for $L=4,6,8$ systems. 
In his results, for small $\gamma$ ($\gamma<\gamma_{c}=2\sim 3$ ) 
there exists a plateau in the magnetization process 
at $M=M_{1/3}\equiv M_{S}/3$ where $M_{S}=3L/2$ is 
the saturation magnetization, 
whereas it seems to disappear for $\gamma >\gamma_{c}$. 
For 3CuCl$_{2}\cdot$2dioxane, 
Hida evaluated the trimer coupling as $\gamma = 4.5$. 
Roji and Miyashita \cite{Roji} calculated the magnetization curve by 
means of quantum Monte Carlo simulation for $\gamma=5$ and a magnetization 
plateau did not appear. 
In these studies, however, 
the phase boundary between the plateau and the no-plateau regions is not 
specified clearly. 

\begin{figure}[h]
\begin{center}
\includegraphics[width=2.2in]{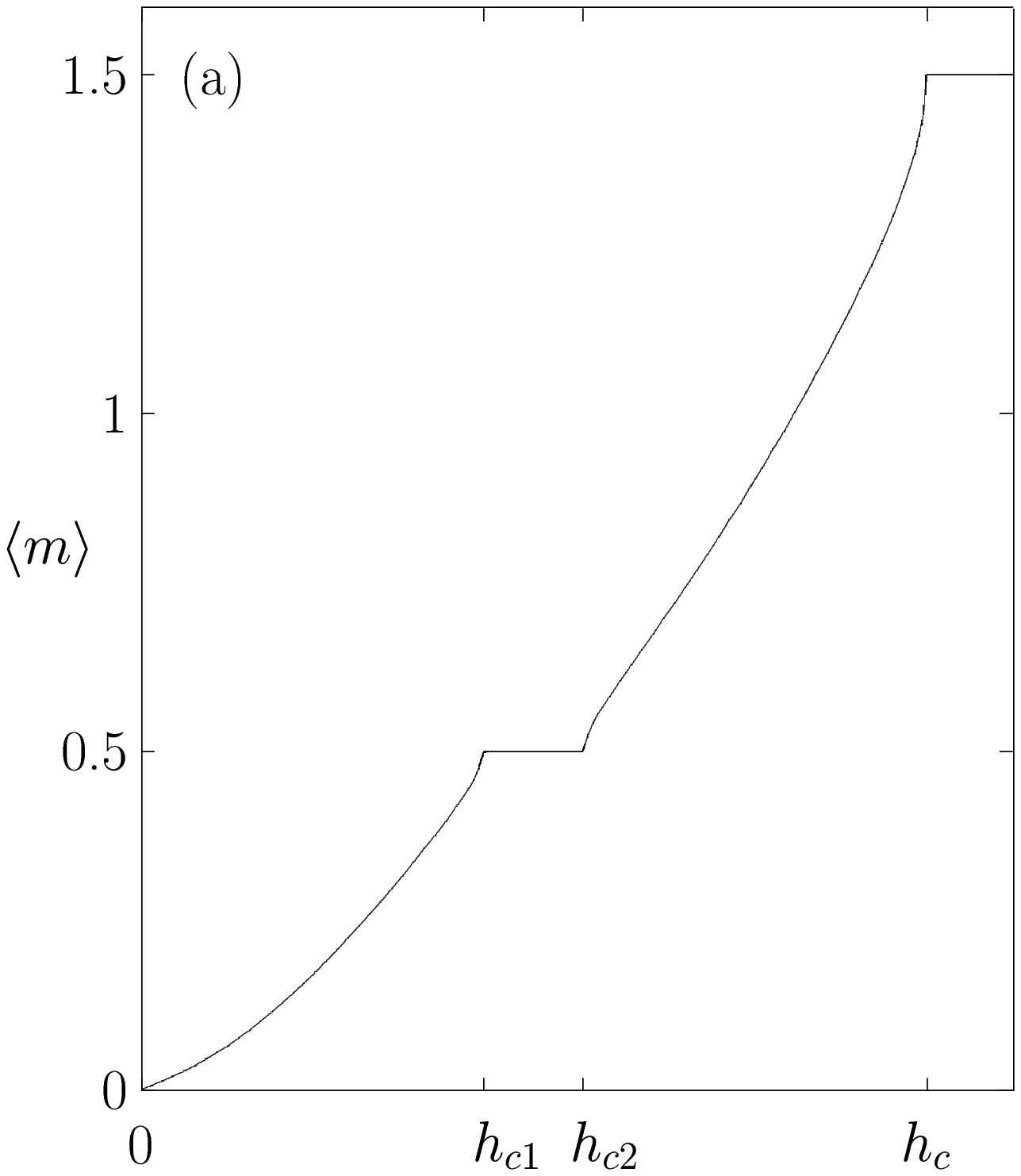}
\hspace{5mm}
\includegraphics[width=2.2in]{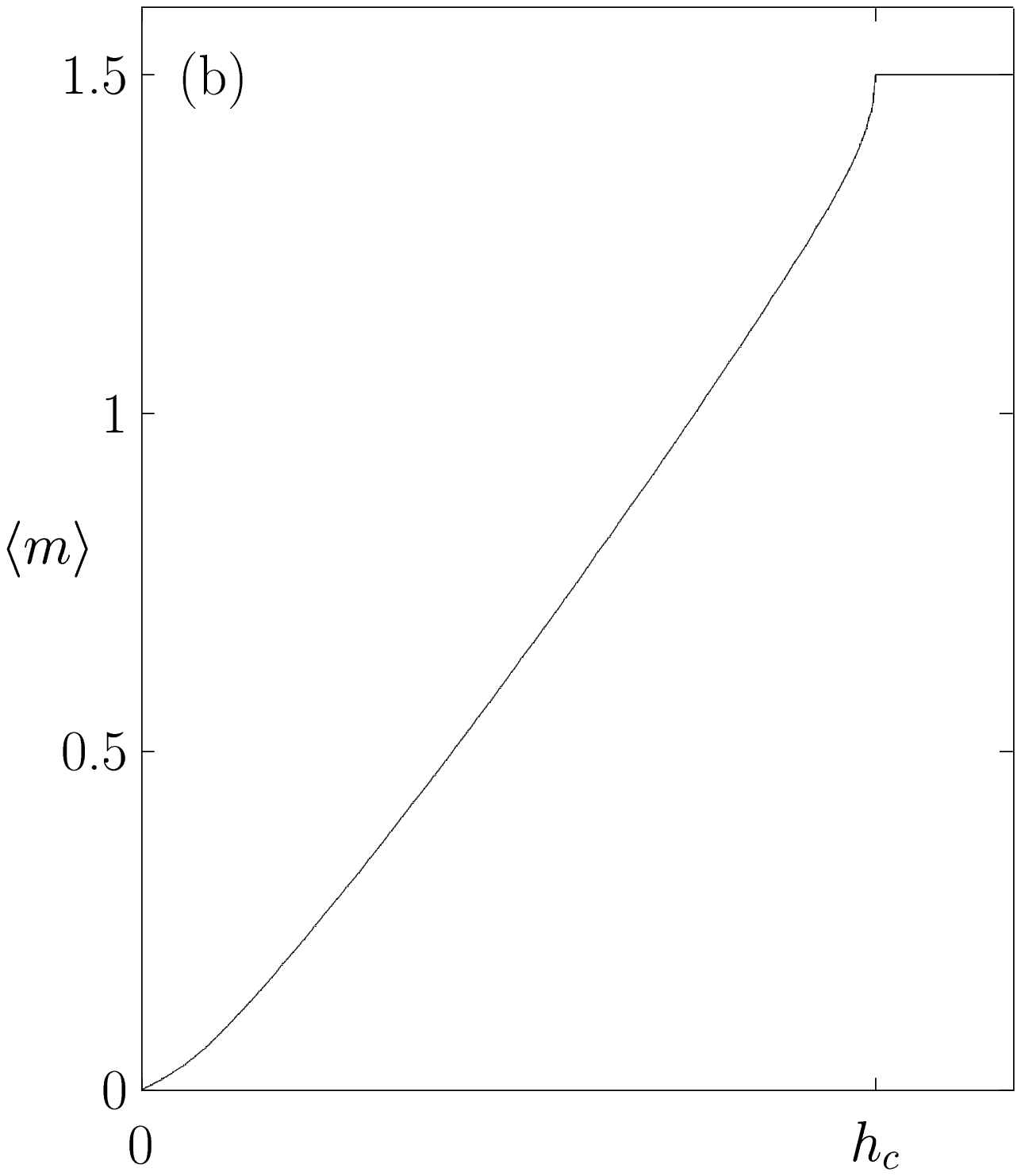}
\end{center}
\caption{Schematic magnetization curve: (a) for $\gamma < \gamma_{c}$; 
(b) for $\gamma > \gamma_{c}$. $h_{c}$ is the saturation field.
$\langle m\rangle$ is the magnetization per trimer $M/L$.}
\end{figure}

For this model, schematic magnetization curves are shown in figure 1. 
For the weak-ferromagnetic-coupling case (figure 1(a)), 
the magnetization plateau appears in the interval $h_{c1}< h < h_{c2}$, 
where the magnetization remains $1/3$ of the saturation magnetization. 
In a crude argument, 
two spins $\mbox{\boldmath$S$}_{3j}$ and $\mbox{\boldmath$S$}_{3j+1}$ 
(connected by antiferromagnetic bond) 
tend to form a singlet, and the remaining $L$ spins are aligned upward. 
In the limit $\gamma\rightarrow \infty$, 
the low-energy behaviour of the present model effectively reduces to that of 
the $S=3/2$ $XXZ$ spin chain \cite{Hida}: 
\begin{equation}
  \tilde{H} = J_{3/2}\sum_{j=1}^{L} \left[
  \tilde{S}_{j}^{x}\tilde{S}_{j+1}^{x}+\tilde{S}_{j}^{y}\tilde{S}_{j+1}^{y}
  +\Delta \tilde{S}_{j}^{z}\tilde{S}_{j+1}^{z}\right]
  -h\sum_{j=1}^{L}\tilde{S}_{j}^{z}
\label{eq:threehalf}
\end{equation}
where $J_{3/2} = J_{AF}/9$ and 
$\tilde{\mbox{\boldmath$S$}}_{j} = \mbox{\boldmath$S$}_{3j-2}+\mbox{\boldmath$S$}_{3j-1} + \mbox{\boldmath$S$}_{3j}$,
and in this limit it is believed 
that there is no magnetization plateau at $M=M_{1/3}$. 
Thus there should be a transition between the plateau and 
the no-plateau (figure 1(b)) regions at the magnetization $M=M_{1/3}$. 
According to the bosonization approach \cite{Okamoto}, this transition 
should be of the Berezinskii-Kosterlitz-Thouless (BKT) type 
\cite{Berezinskii,Kosterlitz1,Kosterlitz2}, 
and the width of the plateau $\Delta h(M=M_{1/3})=h_{c2}-h_{c1}$ 
near the transition point behaves singularly as 
\begin{equation}
  \Delta h \sim \exp (-C/\sqrt{\gamma_{c}-\gamma}) 
\label{eq:pla}
\end{equation}
where $C$ is a constant. 
This singular behaviour makes it difficult to study the critical properties 
numerically. 

In our previous paper \cite{ko-ak}, 
we studied an $S=1/2$ trimerized spin chain: 
\begin{equation}
\fl  H = \sum_{j=1}^{L}\left\{
  (1-t)[H_{3j-2,3j-1}(\Delta) + H_{3j-1,3j}(\Delta)]
  +(1+2t)H_{3j,3j+1}(\Delta) \right\}. 
\label{eq:ham2}
\end{equation}
For this model, we found that in the region surrounded by the three points 
$(\Delta, t) = (-0.729,0)$, $(-1,1)$, and $(-1,-0.5)$, 
there is a no-plateau region at the magnetization $M=M_{S}/3$.  
On the basis of conformal field theory and renormalization group analysis, 
we determined this phase boundary successfully. 
The multicritical point $(\Delta,t)=(-0.729,0)$ is 
consistent with the numerical Bethe-{\it ansatz} calculation: 
$(\Delta, t) = (-0.729043,0)$ \cite{Cabra2}. 
In the present study, 
we apply the same analysis and the numerical approach 
to the model (\ref{eq:ham1}). 

The plan of this paper is as follows. 
In the next section, to study the critical properties, 
we consider an effective continuum model by use of the bosonization technique. 
In section 3, we firstly explain the method used to 
determine the phase boundary between the plateau and the no-plateau regions 
at $M=M_{1/3}$ numerically. 
Next, we show our numerical results for finite-size systems, 
and give the boundary that we obtained 
between the plateau and the no-plateau regions. 
We also check the consistency as regards the critical behaviour. 
The last section is devoted to a summary and discussion. 

\section{Phase transition}
To see the appearance and the disappearance of the magnetization plateau, 
we introduce the effective-continuum model 
of the Hamiltonian (\ref{eq:ham1}). 
In the calculation, it is convenient to use 
the unitary transformation $S_{3j-1}^{x}\rightarrow -S_{3j-1}^{x}$, 
$S_{3j-1}^{y}\rightarrow -S_{3j-1}^{y}$, 
$S_{3j-1}^{z}\rightarrow S_{3j-1}^{z}$ for all $j$, 
and the following parametrization \cite{Okamoto}:
\begin{eqnarray}
\fl  H &=& J_{0}\sum_{j=1}^{L}\left[
    (1+\delta_{\perp})(h_{3j-2,3j-1}^{\perp}+h_{3j-1,3j}^{\perp})
  +(1-2\delta_{\perp})h_{3j,3j+1}^{\perp}\right] \\
\fl  && + J_{0}\sum_{j=1}^{L}\left[
  (\Delta_{0}+\delta_{z})(S_{3j-2}^{z}S_{3j-1}^{z}+S_{3j-1}^{z}S_{3j}^{z})
  + (\Delta_{0}-2\delta_{z})S_{3j}^{z}S_{3j+1}^{z}\right]
  -hM \nonumber
\end{eqnarray}
\begin{equation}
\fl  h_{j,k}^{\perp} = S_{j}^{x}S_{k}^{x} + S_{j}^{y}S_{k}^{y}
\end{equation}
where
\begin{eqnarray}
  J_{0}&=&J_{AF}\frac{2\gamma+1}{3} \nonumber \\
  \delta_{\perp} &=& (\gamma-1)/(2\gamma + 1)  \\
  \Delta_{0} &=& (-2\gamma+\Delta)/(2\gamma + 1) \nonumber \\
  \delta_{z} &=& -(\gamma+\Delta)/(2\gamma+1). \nonumber 
\end{eqnarray}
Using the Jordan-Wigner transformation we firstly map the spin system to 
a spinless-fermion system. Then linearizing the dispersion relation 
of the spinless fermions around the Fermi point, 
and with the standard bosonization procedure, 
we obtain an effective-continuum model 
which describes the low-energy properties of the original system. 

The bosonized Hamiltonian of this model was obtained by 
Okamoto \cite{Okamoto}, as 
\begin{equation}
\fl H = \int \frac{{\rm d}x}{2\pi}\left[ v_{S}K(\pi\Pi)^{2} + \frac{v_{S}}{K}
  \left(\frac{d\phi}{d x}\right)^{2}\right]
  + y_{\phi}v_{S}\int \frac{{\rm d}x}{2\pi}\cos(\sqrt{2}\phi + 2(k_{F}-\pi/3)x)
\label{eq:sg}
\end{equation}
where $v_{S}$ is the sound velocity of the system, 
$\Pi$ is the momentum density conjugate to $\phi$, 
$[\phi(x),\Pi(x')] = i\delta(x-x')$, and 
\[
k_{F} = \frac{\pi}{2}\left(1-\frac{M}{M_{S}}\right)
\]
is the Fermi point of the spinless-fermion model. 
We define the lattice spacing as $1$. 
Due to the oscillation, the second term is important 
only for $M=M_{1/3}$ ({\it i.e.}, $k_{\rm F}=\pi/3$), and  
in the following we consider the system at this magnetization. 
The parameters $v_{S}$, $K$, and $y_{\phi}$ are related to 
$J_{0}$, $\delta_{\perp,z}$ and $\Delta_{0}$ as
\begin{equation}
  v_{S} = J_{0} \sqrt{AC} 
\hspace{5mm} K = \sqrt{\frac{C}{A}} 
\hspace{5mm} y_{\phi}v_{S} = -\pi J_{0}(2\delta_{\perp}+\delta_{z})
\end{equation}
where 
\begin{equation}
  A = \frac{1}{8\pi}\left(1 + \frac{5}{\sqrt{3}\pi}\Delta_{0}\right)
  \hspace{5mm}
  C = 2\pi\left(1-\frac{\Delta_{0}}{\sqrt{3}\pi}\right).
\end{equation}
The dual field $\theta$ conjugate to $\phi$ is defined by 
$\partial_{x}\theta = \pi\Pi$, 
and we make the identification $\phi \equiv \phi+\sqrt{2}\pi$, 
$\theta \equiv \theta+\sqrt{2}\pi$. 
The spin operator in the continuum picture is given by 
\begin{equation}
  S^{z}(x) = \frac{1}{6}+\frac{1}{\sqrt{2}\pi}\frac{d}{dx}\phi(x) 
    +\frac{1}{3}\cos (\sqrt{2}\phi + 2k_{F}x -\pi/3). 
\end{equation}

In the Hamiltonian (\ref{eq:sg}), 
the first term  has a scale invariance and the second 
term violates it. 
For the free-field case $y_{\phi}=0$, the scaling dimension of the operator 
$\sqrt{2}\cos\sqrt{2}\phi$ depends on the Gaussian coupling $K$ as $x=K/2$. 
Thus the second term of equation (\ref{eq:sg}) is relevant or irrelevant 
according to whether the renormalized value of $K$ is 
$K<4$ $(x<2)$ or $K>4$ $(x>2)$. 
In case where $y_{\phi}\neq 0$, 
using the notation $K=4(1+y_{0}/2)$ near $K=4$, 
we have the following renormalization group equations \cite{Kosterlitz2}, 
\begin{equation}
  \frac{{\rm d} y_{0}(L)}{{\rm d}\ln L} = -y_{\phi}(L)^{2} \hspace{5mm}
  \frac{{\rm d} y_{\phi}(L)}{{\rm d}\ln L} = -y_{0}(L)y_{\phi}(L)
\end{equation}
where $L$ is an infrared cut-off. 
For $y_{0} > y_{\phi}>0$, the scaling field $y_{\phi}$ is renormalized to $0$, 
and $y_{0}$ goes to a finite value; 
the system is critical 
and the magnetization plateau does not appear at $M=M_{1/3}$. 
On the other hand, for $y_{0} < y_{\phi}$, 
$y_{\phi}$ flows to infinity and $K$ flows to $0$. 
The phase field is locked to $\phi=\pi/\sqrt{2}$. 
For the spin model (\ref{eq:ham1}), this means that 
the second term in equation (\ref{eq:sg}) makes 
two spins connected by the antiferromagnetic bond form a singlet, 
resulting in the appearance of the magnetization plateau at $M=M_{1/3}$. 
The BKT transition occurs at $y_{0}=y_{\phi}$, 
and at the transition point, we obtain 
\begin{equation}
  y_{0}(L)=\frac{y_{0}}{y_{0}\ln(L/L_{0})+1}, 
\label{eq:logd}
\end{equation}
where $y_{0}(>0)$ and $L_{0}$ are some constants. 
Thus $y_{0}(L)$ flows to $0$ very slowly. 
Near the transition point and $y_{0} < y_{\phi}$, 
the energy gap---that is, the width of the plateau---behaves 
as equation (\ref{eq:pla}) \cite{Kosterlitz2}. 

\section{Numerical method and results}
\subsection{Finite-size behaviour}
In order to study the system numerically, 
let us consider the finite-size behaviour of the model (\ref{eq:sg}). 
The scaling dimension of the primary field 
${\cal{O}}_{m,n}= \exp({\rm i} m\sqrt{2}\phi+{\rm i} n\sqrt{2}\theta)$ for $y_{\phi}=0$
is given by
\begin{equation}
  x_{m,n} = \frac{K}{2}m^{2} + \frac{1}{2K}n^{2}. 
\end{equation}
According to the finite-size scaling theory given by Cardy \cite{Cardy}, 
the excitation energy of the finite-size system at a critical point 
is related to the scaling dimension as 
\begin{equation}
  x_{m,n}(L) = \frac{L}{2\pi v_{S}}(E_{m,n}(L)-E_{g}(L))
\end{equation}
where $E_{g}(L)$ is the ground-state energy of an $L$-site system 
with periodic boundary conditions (PBC). 
Near the BKT transition point ($K\approx 4$), 
the excitation energy is written as 
\begin{equation}
  \frac{L}{2\pi v_{S}}\Delta E_{m,0}(L) = \frac{K(L)}{2}m^{2}
  = 2\left(1+\frac{y_{0}(L)}{2}\right)m^{2}
\label{eq:dimm}
\end{equation}
\begin{equation}
  \frac{L}{2\pi v_{S}}\Delta E_{0,n}(L) = \frac{1}{2K(L)}n^{2}
  = \frac{1}{8}\left(1-\frac{y_{0}(L)}{2}\right)n^{2}
\label{eq:dimn}
\end{equation}
for integer $m$ and $n$. 
In the spinless-fermion language, the integer variable $m$ is 
the difference in number between 
the left- and the right-moving fermions. 
The variable $n$ is the increment of the fermion number, 
and relates to the magnetization as $n=M-M_{1/3}$. 
Thus considering equation (\ref{eq:logd}), 
we have the logarithmic size corrections for finite-size spectrum. 

In the numerical calculation for finite-size systems, 
we calculate the energy of the model (\ref{eq:ham1}), 
fixing the magnetization $M=\sum S_{j}^{z}$ and with $h=0$. 
If the magnetization curve is continuous 
at the magnetization $M=M_{1/3}$, the two values $h_{\pm}(M_{1/3}/L)$: 
\begin{eqnarray}
h_{+}(M_{1/3}/L) &=& E(M_{1/3}+1,L) -E(M_{1/3},L) \nonumber \\
h_{-}(M_{1/3}/L) &=& E(M_{1/3},L) -E(M_{1/3}-1,L) 
\label{eq:numericalM}
\end{eqnarray}
coincide at $L\to\infty$; also,
\[ 
\lim_{L\rightarrow\infty}h_{\pm}(M_{1/3}/L) =h(1/2)
\] 
and this is the magnetic field in the thermodynamic limit. 
Here $E(M,L)$ is the lowest energy for the $L$-site system 
with the magnetization $M$ and $h=0$. 
Then the excitation energy in eq. (\ref{eq:dimn}) can be calculated as
\begin{equation}
  \Delta E_{0,n} = E(M_{1/3}+n,L)  - nh(1/2) - E(M_{1/3},L).
\end{equation}
For the critical system, we have $\Delta E_{0,n}= \Delta E_{0,-n}$. 
Averaging these two values we can eliminate the magnetic filed as
\begin{equation}
\Delta E_{0,\pm n} = \frac{1}{2}[E(M_{1/3}+n,L)+E(M_{1/3}-n,L)]-E(M_{1/3},L). 
\label{eq:theta}
\end{equation}
In the spinless-fermion language, 
the magnetic field plays the role of the chemical potential 
\[
h(M_{1/3}/L)=[E(M_{1/3}+1,L)-E(M_{1/3}-1,L)]/2. 
\]

If the two limits 
\[
\lim_{L\rightarrow\infty}h_{+}(M_{1/3}/L) \hspace{5mm}
\lim_{L\rightarrow\infty}h_{-}(M_{1/3}/L)
\]
do not coincide, there exists a 
magnetization plateau at the magnetization $M_{1/3}$, 
and the width of the plateau is given by 
\[
\Delta h(1/2) = 
\lim_{L\rightarrow\infty}[h_{+}(M_{1/3}/L) - h_{-}(M_{1/3}/L)]
=2\lim_{L\rightarrow\infty}\Delta E_{0,\pm 1}(L).
\]

To determine the BKT transition point, we apply the method 
proposed by Nomura and Kitazawa \cite{Nomura98}, 
in which a level crossing of some excitation levels is used. 
Under the twisted boundary condition (TBC) $S^{x,y}_{3L+1}=-S^{x,y}_{1}$, 
$S^{z}_{3L+1}=S^{z}_{1}$, 
the primary operator ${\cal{O}}_{m,n}$ shifts to the operator 
${\cal{O}}_{m+1/2,n}$ in the quantum spin chain \cite{Alcaraz,Destri,Fukui}. 
For the scaling dimensions of operators 
$\sqrt{2}\cos (\phi/\sqrt{2})$ and $\sqrt{2}\sin (\phi/\sqrt{2})$, 
we have the following finite-size corrections 
\begin{eqnarray}
  x^{c}_{\pm1/2,0}(L) &=& 
  \frac{1}{2}+\frac{1}{4}y_{0}(L)+\frac{1}{2}y_{\phi}(L)
\nonumber \\
  x^{s}_{\pm1/2,0}(L) &=& 
  \frac{1}{2}+\frac{1}{4}y_{0}(L)-\frac{1}{2}y_{\phi}(L).
\label{eq:mag}
\end{eqnarray}
The dependence of the coupling $y_{\phi}$ comes from 
the first-order perturbation of the second term 
in equation (\ref{eq:sg}) \cite{Kitazawa}. 
(Due to the charge neutrality condition in the model (\ref{eq:sg}), 
the operators ${\cal{O}}_{0,n}$ are not corrected in this order.) 

Writing $y_{\phi} = y_{0}(1+t)$ near the transition point, we have 
\begin{eqnarray}
  x^{c}_{\pm 1/2,0}(L)&=&
    \frac{1}{2}+\frac{3}{4}y_{0}(L)\left(1+\frac{2}{3}t\right)
  \nonumber \\
  x^{s}_{\pm 1/2,0}(L)&=&
    \frac{1}{2}-\frac{1}{4}y_{0}(L)(1+2t).
\label{eq:dhalf}
\end{eqnarray}
On the other hand, from equation (\ref{eq:dimn}) 
the scaling dimension of ${\cal{O}}_{0,\pm 2}$ is given by 
\begin{equation}
  x_{0,\pm2}(L) = \frac{1}{2}-\frac{1}{4}y_{0}(L)
  = \frac{L}{2\pi v_{S}}\Delta E_{0,\pm 2}(L). 
\label{eq:ele}
\end{equation}
From these equations, we see that $x_{0,\pm2}$ and $x^{s}_{\pm 1/2,0}$ 
cross linearly at the transition point ($t=0$). 

The energies corresponding to the operators 
$\sqrt{2}\cos (\phi/\sqrt{2})$ and $\sqrt{2}\sin (\phi/\sqrt{2})$ are 
obtained from two low-lying energies with the twisted boundary conditions as 
\begin{eqnarray}
\fl  \Delta E^{c}_{\pm 1/2,0}
  &=& \frac{2\pi v_{S}}{L}x_{\pm 1/2,0}^{c}(L)
  = E^{TBC}\left(M_{1/3},L,P=1\right)
  -E\left(M_{1/3},L\right)
  \nonumber \\
\fl  \Delta E^{s}_{\pm 1/2,0}
  &=& \frac{2\pi v_{S}}{L}x_{\pm 1/2,0}^{s}(L)
  = E^{TBC}\left(M_{1/3},L,P=-1\right)
    -E\left(M_{1/3},L\right).
\label{eq:phi}
\end{eqnarray}
Here $E^{TBC}(M_{1/3},L,P=\pm 1)$ are two low-lying energies under the TBC. 
The two states with these energies 
are distinguished by the parity of the space inversion 
$P: \mbox{\boldmath$S$}_{j}\rightarrow \mbox{\boldmath$S$}_{3L-j+1}$ 
(in the sine-Gordon model (\ref{eq:sg}), $\phi\rightarrow -\phi$). 
Thus the energy differences $\Delta E_{0,\pm2}$ and $\Delta E^{s}_{\pm 1/2,0}$ 
should cross at the transition point. 

\subsection{Results}
In the numerical calculation, we consider finite-size systems 
($L=4,6,8$) with the PBC and the TBC. 
Using the above-mentioned degeneracy at the transition point, 
we determine the phase boundary and also check the universality class
of the phase transition.

For the PBC, the Hamiltonian is invariant under the translation 
$\mbox{\boldmath$S$}_{j}\rightarrow \mbox{\boldmath$S$}_{j+3}$, 
and the corresponding eigenvalue is the wavenumber 
$q=2\pi k/L$ $(k=-L/2+1,\cdots, L/2$). 
In the whole region, the lowest energy states with $M=M_{1/3}, M_{1/3} \pm 2$ 
have the wavenumber $q=0$ and the parity of the space inversion $P=1$. 

\begin{figure}[h]
\begin{center}
\includegraphics[width=3.2in]{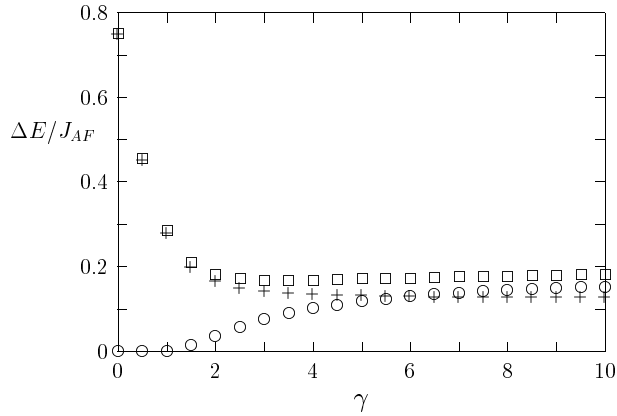}
\end{center}
\caption{Energy differences $\Delta E_{\pm 1/2,0}^{c}$ ($\Box$), 
$\Delta E_{\pm 1/2,0}^{s}$ ($\circ$), and $\Delta E_{0,\pm 2}$ ($+$) 
for the $L=6$, $\Delta=0.5$ system. These values are defined in equations 
(\ref{eq:theta}) and (\ref{eq:phi}) in the text.
We see a level crossing between $\Delta E_{\pm 1/2,0}^{s}$ 
and $\Delta E_{0,\pm 2}$.}
\end{figure}

Figure 2 shows the behaviour of $\Delta E_{0,\pm 2}$ 
and $\Delta E_{\pm 1/2,0}^{c,s}$ for $L=6$ $\Delta = 0.5$ systems. 
There is a level crossing between $\Delta E^{s}_{\pm 1/2,0}$ and 
$\Delta E_{0,\pm 2}$, as is described in equations (\ref{eq:dhalf}), 
and (\ref{eq:ele}). 
The size dependence of the crossing point is shown in figure 3, from which 
the extrapolated value is estimated as $\gamma_{c}(\Delta=0.5)=5.75$. 

\begin{figure}[h]
\begin{center}
\includegraphics[width=3.2in]{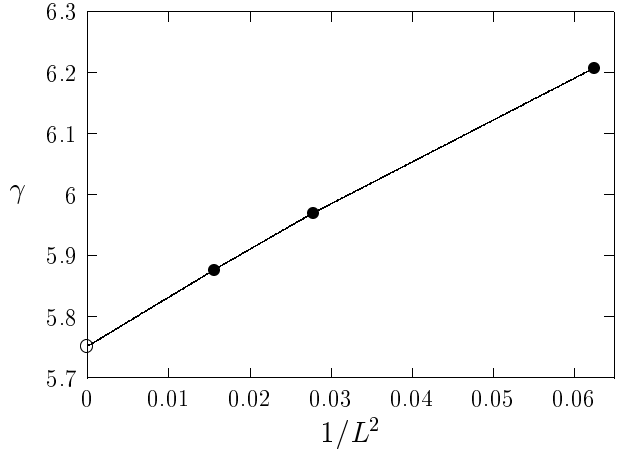}
\end{center}
\caption{The size dependence of the crossing point $\gamma_{c}$ 
for $\Delta=0.5$.}
\end{figure}

In figure 4 we show the crossing points for $L=4, 6, 8$ 
on the $\Delta-\gamma$ plane. 
The line is the extrapolated value 
in the polynomials of $1/L^{2}$, and 
this is the boundary between the plateau and the no-plateau regions. 
This line increases with $\Delta$ from $\gamma =0$ at $\Delta=-1$ 
to $\gamma=15.4$ at $\Delta=1$. 
This is explained by the fact that 
when we decrease $\Delta$ from $1$, the dimerization effect 
in the antiferromagnetic bond is weakened. 

\begin{figure}[h]
\begin{center}
\includegraphics[width=3.2in]{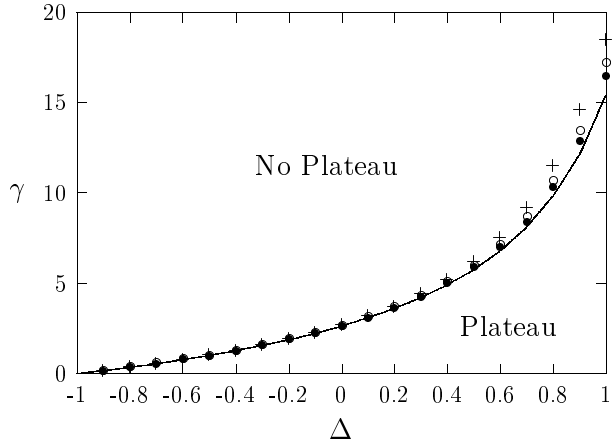}
\end{center}
\caption{Crossing points for $L=4$ ($+$), $L=6$ ($\circ$), 
and $L=8$ ($\bullet$). The solid line shows the extrapolated values.}
\end{figure}

As a consistency check with the critical theory, 
we calculate the following averaged scaling dimension: 
\begin{equation}
  \frac{x_{\pm 1/2,0}^{c}(L) + 3x_{\pm 1/2,0}^{s}(L)}{4}.
\label{eq:average}
\end{equation}
Taking this average, we can eliminate 
the leading logarithmic correction (\ref{eq:logd}) 
for finite-size systems at the critical point $t=0$ 
(see equation (\ref{eq:dhalf})). 
To calculate the scaling dimension, we need the sound velocity $v_{S}$. 
This can be calculated using the lowest energy with 
$M=M_{1/3}$ and the wavenumber $q=2\pi/L$ 
(corresponding to the U(1) current): 
\begin{equation}
  v_{S} = \lim_{L\rightarrow \infty} 
  \frac{E(M_{S}/3,L,q=2\pi/L)-E(M_{S}/3,L)}{2\pi/L}.
\label{eq:velocity}
\end{equation}
Figure 5 shows the extrapolated value 
of the averaged scaling dimension (\ref{eq:average}) 
and the bare value of the $L=8$ system on the transition line. 
The averaged scaling dimension is very close to the expected value $x=1/2$, 
whereas the bare values $x_{\pm 1/2,0}^{c}(L)$ and $x_{\pm 1/2,0}^{s}(L)$ 
are far from $x=1/2$ due to the logarithmic size correction. 

To confirm the universality class of the phase transition 
for the $M=M_{1/3}$ systems, 
we also calculate the conformal anomaly number $c$ 
from \cite{Blote,Affleck}
\begin{equation}
  \frac{E(M_{S}/3,L)}{L} = \epsilon(M_{S}/3) -\frac{\pi v_{S}c}{6L^{2}}
  + \cdots. 
\end{equation}
For the BKT critical point, the conformal anomaly number is $c=1$. 
We see that the value $c$ is close to $1$ 
within the error of a few percent ({\it e.g.} $c=1.01$ for $\Delta=1$). 
Thus we can conclude that the transition 
between the plateau and the no-plateau 
region at $M=M_{1/3}$ is of the BKT type.  

\begin{figure}[h]
\begin{center}
\includegraphics[width=3.2in]{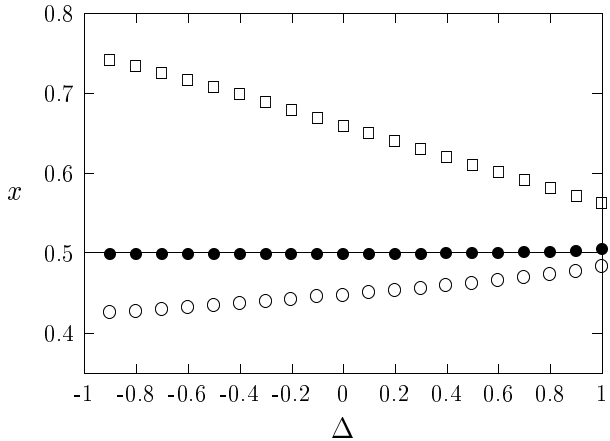}
\end{center}
\caption{Extrapolated values of the averaged scaling dimension 
$(x_{\pm 1/2,0}^{c}+3x_{\pm 1/2,0}^{s})/4$ on the transition line ($\bullet$). 
We also show the bare values of $x_{\pm 1/2,0}^{c}(L=6,\Box)$ 
and $x_{\pm 1/2,0}^{s}(L=6,\circ)$.}
\end{figure}

\section{Summary and discussion}
We studied the plateau--no-plateau transition of the model 
(\ref{eq:ham1}) for the magnetization $M=M_{S}/3$. 
For small $\gamma$ there is a plateau in the magnetization process, 
while for large $\gamma$ there is no plateau (figure 1). 
By use of the effective-continuum model and the renormalization group 
method, we showed that this quantum phase transition is 
of the Berezinskii-Kosterlitz-Thouless type. 
On the basis of conformal field theory and renormalization group analysis, 
we argued for the finite-size spectrum and 
determined the phase boundary (figure 4) 
using a level crossing of special excitations (figure 2). 
This boundary runs from ($\Delta,\gamma)=(-1,0)$ 
to $(\Delta,\gamma)=(1,15.4)$. 
We also checked the consistency of the critical theory 
and the numerical calculation, and concluded that 
the phase transition is of the BKT type. 

The extrapolated critical point for $\Delta=1$ is $\gamma = 15.4$, which 
is somewhat larger than the value from previous experimental and numerical 
studies \cite{Hida,Ajiro,Roji}. 
One possible explanation for this discrepancy is as follows. 
The width of the plateau near the transition point is given by 
equation (\ref{eq:pla}), and the coefficient in 
exponential, $C$, is large for $\Delta=1$. 
Thus the narrow-plateau-width region (that is, the small-gap region) 
is widely ranged. 
This means that 
it is difficult to estimate the plateau width $\Delta h$ directly 
from the numerical calculation and also at finite temperatures. 
According to Kosterlitz \cite{Kosterlitz2}, 
the coefficient $C$ is described as $C \propto 1/\sqrt{y_{0}}$, 
and from figure 5, we see that $y_{0}(=y_{\phi})$ 
is small at the transition point for $\Delta=1$. 
For the $S=3/2$ antiferromagnetic Heisenberg chain 
corresponding to the large-$\gamma$ limit, 
the Gaussian coupling $K$ is calculated as $K\approx 4.4$
(the compactification radius is $R=(2\pi K)^{-1/2}=0.95/\sqrt{8\pi}$) 
in reference \cite{Oshikawa}, 
which is slightly larger than the critical value $K=4$. 
This fact supports a large value of $\gamma_{c}$ for $\Delta=1$. 

\ack
We thank K. Nomura and M. Oshikawa for useful discussions. 
The computation for this work was done partially using the facilities 
of the Supercomputer Center, Institute for Solid State Physics, 
University of Tokyo.

\end{document}